\documentclass{article}
\usepackage{hyperref,spconf,amssymb,amsmath,graphicx,ctable,makecell,multirow,subcaption}
\usepackage[super]{nth}
\usepackage{enumitem}

\title{Effective Low-Cost Time-Domain Audio Separation Using\\Globally Attentive Locally Recurrent Networks}
\name{Max W. Y. Lam$^1$, Jun Wang$^1$, Dan Su$^1$, Dong Yu$^2$}
\address{
  $^1$Tencent AI Lab, Shenzhen, China\\
  $^2$Tencent AI Lab, Bellevue WA, USA}
\begin{document}
\ninept
\maketitle
\begin{abstract}
Recent research on the time-domain audio separation networks (TasNets) has brought great success to speech separation. Nevertheless, conventional TasNets struggle to satisfy the memory and latency constraints in industrial applications. In this regard, we design a low-cost high-performance architecture, namely, globally attentive locally recurrent (GALR) network. Alike the dual-path RNN (DPRNN), we first split a feature sequence into 2D segments and then process the sequence along both the intra- and inter-segment dimensions. Our main innovation lies in that, on top of features recurrently processed along the inter-segment dimensions, GALR applies a self-attention mechanism to the sequence along the inter-segment dimension, which aggregates context-aware information and also enables parallelization. Our experiments suggest that GALR is a notably more effective network than the prior work. On one hand, with only $1.5$M parameters, it has achieved comparable separation performance at a much lower cost with $36.1\%$ less runtime memory and $49.4\%$ fewer computational operations, relative to the DPRNN. On the other hand, in a comparable model size with DPRNN, GALR has consistently outperformed DPRNN in three datasets, in particular, with a substantial margin of $2.4$dB absolute improvement of SI-SNRi in the benchmark WSJ0-2mix task.
\end{abstract}

\section{Introduction}

Audio separation is a fundamental problem in signal processing, and the most typical problem is called ``cocktail party problem'' \cite{cherry1953some} including multi-talker speech separation, overlapped speech-music separation, etc.
Recent advances in deep learning models \cite{luo2019dual, liu2019divide, morrone2019face, yu2020audio} have drastically advanced state-of-the-art speech separation performances on several benchmark datasets.
Currently, one outstanding category of the best-performing solutions is based on the time-domain audio separation network (TasNet) \cite{luo2018tasnet}, which takes mixture waveforms as inputs and directly reconstruct sources by computing time-domain loss with permutation invariant training (PIT) \cite{yu2017permutation, kolbaek2017multitalker}. In particular, there were several types of TasNets: the initially proposed bi-directional long short-term memory (Bi-LSTM) based TasNet \cite{luo2018tasnet}, the time convolutional network (TCN) based Conv-TasNet \cite{bai2018empirical, luo2019conv}, the dual-path recurrent neural network (DPRNN) \cite{luo2019dual} and the recently proposed dual-path Transformer network (DPTNet) \cite{chen2020dual}.

\begin{figure}[thb]
\vspace{-0.1cm}
 \centering
    \includegraphics[width=0.48\textwidth]{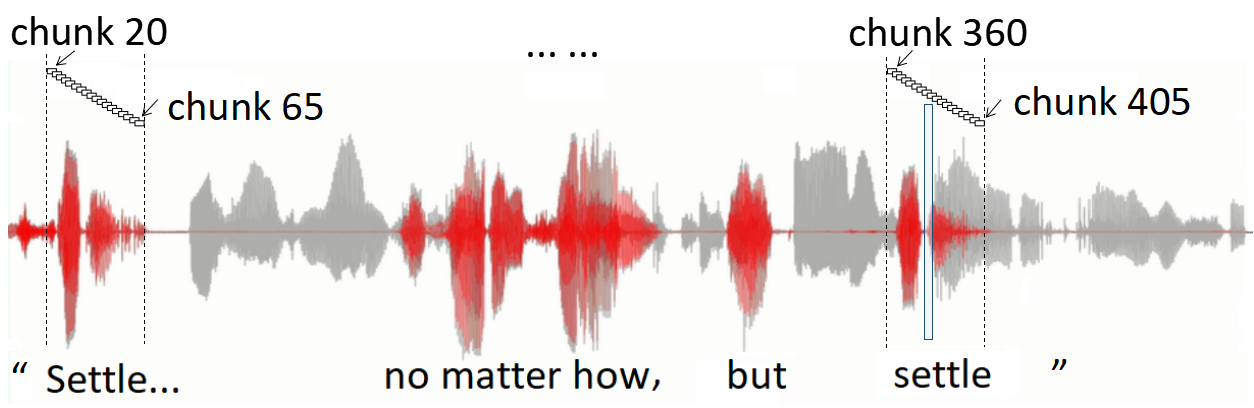}
    \includegraphics[width=0.4\textwidth]{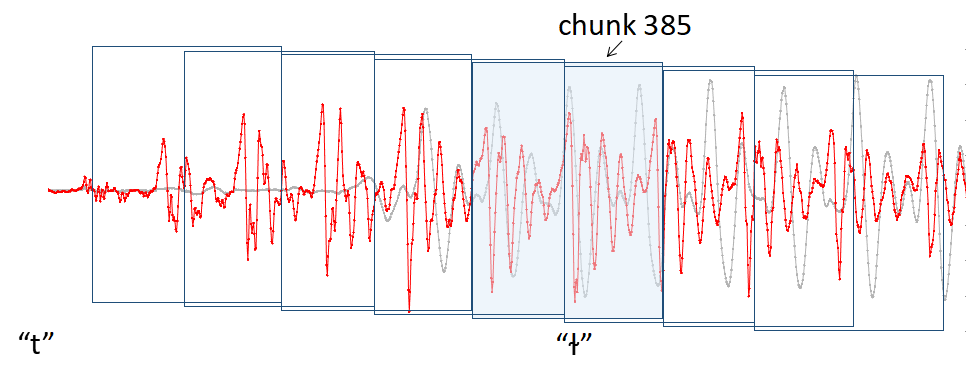}
\caption{Upper: a 4s raw waveform mixture of two overlapping utterances; Lower: zooming in on the 385th segment around the lateral phoneme of /l/.}
\vspace{-0.5cm}
\label{fig:4s}
\end{figure}

These TasNet-based prior work has proven that a smaller window size improves the separation performance at the cost of a significantly longer 1-D feature sequence \cite{luo2019dual,chen2020dual}. To provide a more concrete illustration, we take a $4$-second $16$Hz sample rate waveform input as an example in Fig. \ref{fig:4s}, where the resultant feature sequence that a TasNet (with a window size of $2$ samples and hop size of $1$ sample) needs to model would be as long as $64000$. Learning such long-term sequential dependency poses special challenges to various conventional sequential modeling networks, including attention models\cite{giri2019attention,kim2019transformer}, CNNs \cite{daniel2018unet,luo2019conv}, and RNNs (e.g., LSTMs\cite{gers1999learning} and GRUs \cite{chung2014empirical}), each with respective difficulties as discussed below.

Attention models \cite{vaswani2017attention} has superiority in learning context-aware long-term dependencies, e.g., in BERT \cite{devlin2018bert} for natural language processing tasks. Most recently, a series of research has also attempted to apply self-attention in speech signal processing, but generally to remarkably shorter feature sequences than a raw input, e.g., frame-level acoustic features for speech recognition \cite{dong2018speech, pham2019very, han2019multi, luo2020simplified}, layer features in a U-Net architecture \cite{giri2019attention} or short-time Fourier transform (STFT) features for speech enhancement \cite{kim2019transformer}. Nevertheless, attention models have been hardly applied to time-domain source separation tasks as we discussed above because its complexity and memory consumption per layer is quadratic to the sequence length and become unacceptable for very long sequential modeling.
1-D CNNs with fixed receptive fields that are smaller than the very long sequence length, unlike RNNs that have dynamic receptive fields, are not able to fully utilize the sequence-level dependency \cite{bai2018empirical}.
RNNs are also limited by its nature of recursively processing and memorizing context \cite{khandelwal2018sharp,ravanelli2018light,wang2019r}. To mitigate the long sequence modeling problem for RNNs, Luo et al. \cite{luo2019dual} introduced DPRNN, in which the long signal sequence is divided into shorter segments and interleave two RNNs, an inter-segment RNN and an inter-segment RNN, for local and global modeling, respectively.

To provide a better panorama about how a sequential context a TasNet (e.g., DPRNN) is dealing with looks, we plot the 4s raw waveform mixture in the upper in Fig. \ref{fig:4s}, where one of the two overlapping utterances, saying ``Settle, no matter how, but settle'', has been marked in red. Under the dual-path setting, it is segmented into 512 segments, each with length 250. The lower plot zooms in on the \nth{385} segment to show the details inside a segment around the lateral phoneme of /l/ in the second ``settle''. 
As we can see, high temporal correlation, acoustic signal structure, and continuities occur in inter-segment sequences, whilst strong discontinuities occur in inter-segment sequences. As revealed by Khandelwal et al. \cite{khandelwal2018sharp}, RNNs are far more sensitive to the nearby elements than to the distant ones, and the model is capable of using about $200$ tokens of context on average, but sharply distinguishes nearby context (recent $50$ tokens) from the distant history. This suggests that RNNs are ideally suited for inter-segment modeling, but not necessarily so for inter-segment modeling. Moreover, Ravanelli et al. \cite{ravanelli2018light} discovered that RNNs reset the stored memory to avoid bias towards an ``unrelated'' history. However, unlike language modeling where information could be safely discarded when moving from one text to another semantically unrelated text, we argue that for specific tasks such as audio separation, faraway memory could potentially be important. For example, as shown in Fig. \ref{fig:4s}, the first ``settle'' (\nth{20}- \nth{65} segment), which is very distant from the second ``settle'' (\nth{360}- \nth{405} segment), could be more useful than nearby elements. In contrast, one strength of attention mechanisms over RNNs lies in a fully connected sequence processing strategy, where every element is connected to other elements in a sequence via a direct path without any recursively processing, memorizing reset, or update mechanisms like RNNs. Given the above example, for the second ``settle'', a self-attention model would have readily placed more importance to the first ``settle'', despite the faraway context beyond $200$ segments.

Motivated by the above observation, our work revises TasNets and DPRNN to better address long-range context modeling for audio separation, leading to a lower-cost, higher-performance structure called globally attentive locally recurrent (GALR) network. We resort to the self-attention mechanism \cite{vaswani2017attention} for inter-segment computation to model context-aware global dependencies. Meanwhile, we keep using RNNs for modeling local dependencies at the lower inter-segment context level, e.g., signal continuity, signal structure, etc, which are inherently important for waveform reconstruction.

The contribution of this paper is four-fold:
\begin{itemize}[leftmargin=*]
\item To the best of our knowledge, this is the first work that jointly leverages the attention and recurrent mechanisms in an alternate and iterative manner, and most importantly, allows the system to take advantage of both techniques, which are complementary at modeling global long-range context and local detail dependencies, respectively. 

\item Our work elegantly solves the critical bottleneck of self-attention networks due to unacceptable computational and memory cost for modeling very long sequences, as we can control the global sequence length via the dual-path structure \cite{luo2019dual}. Related similar work includes R-Transformer \cite{wang2019r} and DPTNet \cite{chen2020dual}. Nevertheless, R-Transformer has the attention mechanism at the middle-level directly applied to the whole sequence, thus its computational cost is still quadratic to the sequence length, which hinders its application to TasNet; DPTNet also applied a dual-path setting like DPRNN to a Transformer-based architecture, but both inter-segment and inter-segment sequences are processed by a combination of attention model and RNN, leading to additional computational cost much heavier than DPRNN and our proposed GALR, which we will discuss in more details in Section \ref{sec:4.2.3}.

\item The proposed GALR model has a significantly lower cost with 42.3\% reduction in model-size and with a 36.1\% reduction in runtime memory cost and a 49.4\% reduction in computational operations, while achieving comparable or better performance comparing to DPRNN.
Moreover, unlike RNNs and DPRNN, the global attentive structure can further reduce training and inference time via parallelization, as the attention can be computed for all segments in parallel and allows parallel computation for aggregating information across segments over long audio sequences.

\item Finally, results show consistently higher performance over DPRNN in terms of SI-SNRi and SDRi across three different datasets, while still maintaining a lower cost.
\end{itemize}



\begin{figure*}[ht!]
\vspace{-0.1cm}
 \centering
    \includegraphics[width=0.8\linewidth]{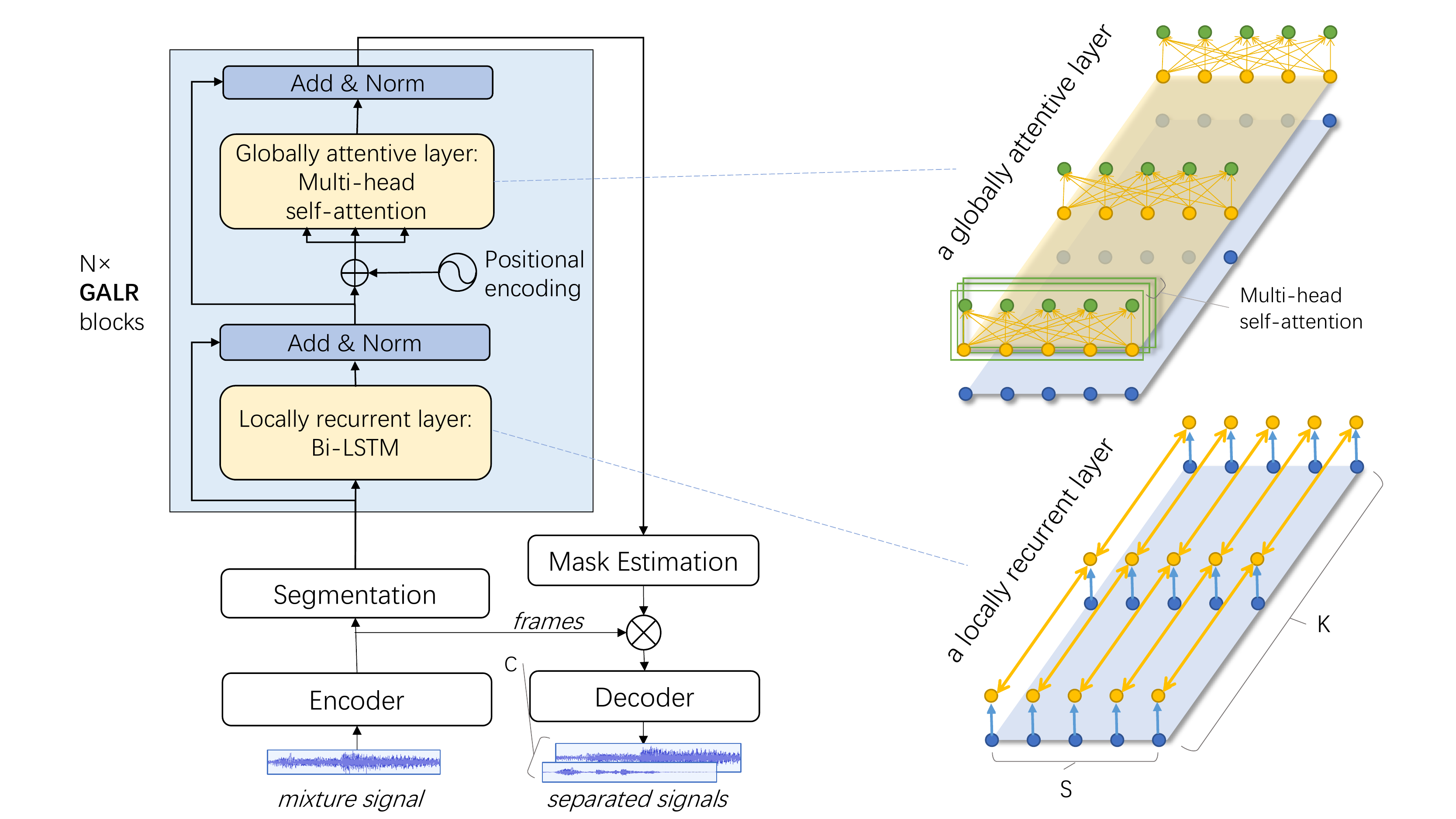}
\caption{Left: the overall architecture of our GALR network. Right: detailed illustration about how the intra- and inter-segment sequences are processed in the locally recurrent layer (lower right) and the globally attentive layer (upper right) inside each GALR block, respectively.}
\vspace{-0.3cm}
\label{fig:archi}
\end{figure*}

\section{Model Design}
\label{sec:2}
This section presents our proposed globally attentive locally recurrent (GALR) network. Fig. \ref{fig:archi} shows the inner machinery of GALR, of which the core processing component is a stack of GALR blocks. Each GALR block contains two modeling perspectives. The first modeling perspective is responsible for modeling the local structures of input signals recurrently; the second modeling perspective aims at capturing global dependencies with the multi-head self-attention mechanism. Next, we describe each part in detail.

\subsection{Encoding Raw Signals}

\subsubsection{Encoder}
In a TasNet-based separation system, an input mixture signal is represented as $I$ half-overlapping frames, denoted by $\mathbf{x}_1, ..., \mathbf{x}_I\in\mathbb{R}^{M}$, where $M$ denotes the window length. Analogous to the short-time Fourier transform (STFT), we non-linearly transform each frame $\mathbf{x}_i$ into a $D$-dimensional feature vector $\mathbf{\Tilde{x}}_i\in \mathbb{R}^{D}$ using a 1D gated convolutional layer:
\begin{align}
    \mathbf{\Tilde{x}}_i=\text{ReLU}(\mathbf{U}*\mathbf{x}_i),
\end{align}
where $*$ denotes the 1D convolution operation, $\mathbf{U}\in \mathbb{R}^{D\times M}$ contains $D$ vectors (encoder basis functions) with length $M$ each, and $\text{ReLU}(\cdot)$ is the rectified linear unit used in \cite{luo2018tasnet, luo2019dual, luo2019conv} to ensure the non-negativity.

\subsubsection{Segmentation}
Given an encoded signal input $\mathbf{\Tilde{X}}\in \mathbb{R}^{D\times I}$, the segmentation module splits $\mathbf{\Tilde{X}}$ into $S$ half-overlapping segments each of length $K$. The first and last segments are padded with zeros to create $S=\lceil 2I/K \rceil +1$ equal-size segments: $\mathbf{S}_s\in\mathbb{R}^{D\times K}$ for $s=1,...,S$. These segments are packed into a 3D tensor, denoted by $\mathbf{T} \in \mathbb{R}^{D\times S \times K}$. Note that the size of each segment $K$ is a hyperparameter that affects the number of segments and can be used to control the scale of the locality.

\subsection{GALR Blocks}
The input 3D tensor $\mathbf{T}$ is then passed to a stack of $N$ GALR blocks, as shown in Fig. \ref{fig:archi}, which is designed to decouple the mixture signal by alternating local and global sequence modeling. Each GALR block returns a 3D tensor with the same dimensionality as its input. We denote the input for block $n = 1, ..., N$ as $\mathbf{T}^{(n)} \in \mathbb{R}^{D\times S \times K}$, where $\mathbf{T}^{(1)}=\mathbf{T}$. As shown on the right of Fig. \ref{fig:archi}, a GALR block is composed of two phases of computation, a locally recurrent layer and a globally attentive layer, respectively corresponding to inter-segment processing and inter-segment processing. Each of which is described in detail below.

\subsubsection{Locally Recurrent Model}
A recurrent model is adopted to model the local information of the input sequence upon segmentation. To model such short-term dependencies within each segment, we employ a bi-directional LSTM of $H$ hidden nodes:
\begin{align}
    \mathbf{L}^{(n)} = \left[\mathbf{R}^{(n)}f^{(n)}_\text{Bi-LSTM}\left(\mathbf{T}^{(n)}[:, s, :]\right)+\mathbf{Y}^{(n)}, s = 1, . . . , S\right],
\end{align}
where $\mathbf{R}^{(n)}\in \mathbb{R}^{D \times 2H}$ and $\mathbf{Y}^{(n)}\in \mathbb{R}^{D \times 1}$ form a linear layer, $\mathbf{L}^{(n)}\in \mathbb{R}^{D\times S \times K}$ is the output of the Bi-LSTM $f^{(n)}_\text{Bi-LSTM}(\cdot)$, and $\mathbf{T}^{(n)}[:, s, :]\in \mathbb{R}^{D\times K}$ refers to the local sequence within the $s^\text{th}$ segment. The output of this locally recurrent model then goes through a layer normalization operation $\text{LN}(\cdot)$ with a residual connection to the block's input, which in practice is critical for model regularization and training acceleration \cite{luo2019dual}:
\begin{align}
    \mathbf{\hat{L}}^{(n)} = \text{LN}(\mathbf{{L}}^{(n)})+ \mathbf{T}^{(n)}.
\end{align}
\subsubsection{Globally Attentive Model}
We build a globally attentive model on top of the locally recurrent model to capture the long-term dependencies. Recent works in the speech community \cite{dong2018speech, li2019neural, ren2019fastspeech} have unveiled the extraordinary performance of attention mechanisms in learning long-term global dependencies \cite{vaswani2017attention}. Here in our case, the multi-head attention mechanism especially becomes the perfect fit due to three reasons. Firstly, the inherent computational burden of attention models becomes less-problematic since we now can control the sequence length by changing the window length of segmentation. Secondly, global dependencies across segments are directly modeled without needing to memorize segments one by one as in RNNs. Thirdly, given that the input is composed of different sources, it is sensible to use multiple attention schemes (a.k.a. heads) on the whole sequence.
\par
Before applying the attention mechanism, the output of locally recurrent model first goes through the following:
\begin{align}
\label{eq:3}
    \mathbf{G}^{(n)} = \text{LN}_{D}(\mathbf{\hat{L}}^{(n)})+\mathbf{P}
\end{align}
where $\text{LN}_{D}(\cdot)$ denotes the layer normalization applied only along the feature dimension $D$, $\mathbf{P}$ denotes the positional encoding matrix introduced in \cite{vaswani2017attention}. For global sequence modeling, we consider the sequence of frames across all segments, i.e.,  $\mathbf{G}^{(n)}_k \triangleq [\mathbf{G}^{(n)}[:, s, k], s=1,...,S]$. In order to create $J$ heads, we linearly map $\mathbf{G}^{(n)}_k$ into $J$ different forms of query, key, and value matrices:
\begin{align}
\mathbf{Q}^{(n)}_{k,j} &= \mathbf{W}^{(n)}_{\text{query},j}(\mathbf{W}^{(n)}_\text{query}\mathbf{G}^{(n)}_k+\mathbf{b}^{(n)}_\text{query})\\
\mathbf{K}^{(n)}_{k,j} &= \mathbf{W}^{(n)}_{\text{key},j}(\mathbf{W}^{(n)}_\text{key}\mathbf{G}^{(n)}_k+\mathbf{b}^{(n)}_\text{key})\\
\mathbf{V}^{(n)}_{k,j} &= \mathbf{W}^{(n)}_{\text{value},j}(\mathbf{W}^{(n)}_\text{value}\mathbf{G}^{(n)}_k+\mathbf{b}^{(n)}_\text{value})
\end{align}
for $k=1,...,K$, $j=1,...,J$.
, where $\mathbf{W}^{(n)}_\text{query}$, $\mathbf{W}^{(n)}_\text{key}$, $\mathbf{W}^{(n)}_\text{value} \in \mathbb{R}^{D\times D}$, $\mathbf{b}^{(n)}_\text{query}$, $\mathbf{b}^{(n)}_\text{key}$, $\mathbf{b}^{(n)}_\text{value} \in \mathbb{R}^{D\times 1}$ and $\mathbf{W}^{(n)}_{\text{query},j}$, $\mathbf{W}^{(n)}_{\text{key},j}$, $\mathbf{W}^{(n)}_{\text{value},j} \in \mathbb{R}^{D/J\times D}$.
Note that the attention parameters are not dependent on $k$, which means that we tie the attention weights for all $K$ sequences, i.e, $[\mathbf{G}^{(n)}_k, k=1,...,K]$. Tying weights is sensible here because the cross-segment sequences formed within a relatively small segment size ought to have very high correlations. 

Given the query, key, and value inputs, we then compute the scaled dot-product attention following \cite{vaswani2017attention} in $J$ heads:
\begin{align}
\label{eq:5}
\mathbf{A}^{(n)}_{k,j} = \text{Softmax}\left(\frac{{\mathbf{Q}^{(n)}_{k,j}}^\top\mathbf{K}^{(n)}_{k,j}}{\sqrt{D/J}}\right)\mathbf{V}^{(n)}_{k,j}.
\end{align}

Next, the attention matrices computed at different heads are combined using an affine transformation after concatenating the matrices:
\begin{align}
\label{eq:8}
\mathbf{A}^{(n)}_k = \mathbf{W}^{(n)}_\text{attn}\text{Concat}\left(\mathbf{A}^{(n)}_{k, 1},...,\mathbf{A}^{(n)}_{k, J}\right),
\end{align}
where $\mathbf{W}^{(n)}_\text{attn}\in\mathbb{R}^{D\times D}$ is the weight matrix for the heads. The attention outputs are then concatenated back to a 3D tensor, i.e,
$\mathbf{A}^{(n)} = [\mathbf{A}^{(n)}_k, k = 1, . . . , K]$. Given the attention output, we employ a sub-layer connection with reference to the well-known Transformer model \cite{vaswani2017attention} given by
\begin{align}
    \mathbf{\hat{G}}^{(n)}=\text{LN}(\mathbf{G}^{(n)}+\text{Dropout}(\mathbf{A}^{(n)})),
\end{align}
where $\text{Dropout}(\cdot)$ denotes the dropout regularization \cite{srivastava2014dropout} operation. Finally, the GALR block outputs a residual sum between the local model output and global model output:
\begin{align}
    \label{eq:10}
    \mathbf{T}^{(n+1)} = \mathbf{\hat{G}}^{(n)}+ \mathbf{\hat{L}}^{(n)},
\end{align}
which defines a recurrence relation between $N$ GALR blocks.

\subsubsection{Low-dimension Segment Representation}
Note that the attention mechanism is repeated $K$ times in Eq. (\ref{eq:5}-\ref{eq:8}), we observe that our globally attentive model can use a low-dimension trick to reduce memory and floating-point operations, while maintaining the performance. Due to high correlations between the cross-segment sequences, we indeed can approximate the global dependencies with a down-sampled number of sequences. In this regard, we employ two affine transformations $C_\text{map}(\cdot)$ and ${C}_\text{inv}(\cdot)$ for mapping $K$ dimensions into $Q$ dimensions and inversely mapping $Q$ dimensions back to $K$ dimensions, respectively, where $Q << K$. Mathematically, we only need to re-write Eq. (\ref{eq:3}) and Eq. (\ref{eq:10}) as
\begin{align}
    \mathbf{G}^{(n)} = \text{LN}_{D}({C}_\text{map}(\mathbf{\hat{L}}^{(n)}))+\mathbf{P},\\
    \mathbf{T}^{(n+1)} = {C}_\text{inv}(\mathbf{G}^{(n)})+\mathbf{\hat{L}}^{(n)},
\end{align}
where $C_\text{map}(\cdot)$ and ${C}_\text{inv}(\cdot)$ contain affine mapping parameters in shapes $Q\times (K+1)$ and $K\times {(Q+1)}$, respectively.

\subsection{Signals Reconstruction}
\subsubsection{Mask Estimation}
After $N$ consecutive GALR blocks, we obtain a representation of the mixture signal that facilitates the separation of $C$ sources. We then use a 2D convolutional layer to transform this 3D representation into $C$ 3D tensors. Then, we transform each of the $C$ 3D tensors back to a matrix $\mathbf{S}_c\in\mathbb{R}^{D\times L}$ for $c=1,...,C$ by applying the OverlapAdd method described in \cite{luo2019dual}. After that, we have a beam-forming procedure \cite{luo2019fasnet} that applies two 1D gated convolution layers to each of the $C$ matrices:
\begin{align}
    \mathbf{\hat{S}}_c = \tanh(\mathbf{U}_\text{tanh}*{\mathbf{S}_c})\odot\sigma(\mathbf{U}_\text{sigmoid}*{\mathbf{S}_c}),
\end{align}
where $\odot$ denotes element-wise multiplication, $\sigma(\cdot)$ is the Sigmoid function, and $\mathbf{U}_\text{tanh}\in\mathbb{R}^{D\times D}$ and $\mathbf{U}_\text{sigmoid}\in\mathbb{R}^{D\times D}$ are two parameter matrices in the 1D gated convolution. The Tanh and Sigmoid functions here act as the beam-forming filters.
\par
To produce a mask matrix for each source, the final step is to employ a resilient linear (ReLU) mask function
\begin{align}
    \mathbf{M}_c = \text{ReLU}(\mathbf{U}_\text{relu}*\mathbf{\hat{S}}_c),
\end{align}
where $\mathbf{U}_\text{relu}\in\mathbb{R}^{D\times D}$ is a 1D convolution for learning mask.

\subsubsection{Decoder for Waveform Reconstruction}
The $c^{th}$ estimated mask is applied back to the initially encoded mixture $\Tilde{\mathbf{X}}=[\Tilde{\mathbf{x}}_1,...,\Tilde{\mathbf{x}}_L]$ to reconstruct source $c$:
\begin{align}
    \hat{\mathbf{s}}_c = \text{OverlapAdd}(\mathbf{B}(\Tilde{\mathbf{X}}\odot \mathbf{M}_c)),
\end{align}
where $\mathbf{B}\in\mathbb{R}^{M\times D}$ is a matrix containing the basis signals with each column corresponding to a 1D filter.

\begin{table*}[t]
\centering
\caption{Comparison of dual-path processing time complexities for different block types.}
\vspace{-0.1cm}
\label{tab:0}
\begin{tabular}{cccc}
\specialrule{.16em}{0em}{0em} 
\textbf{Block} & \textbf{Local-path Complexity} & \textbf{Global-path Complexity} & \textbf{Maximum Path Length}
\cite{vaswani2017attention}\\
\hline
DPTNet \cite{chen2020dual} & $\mathcal{O}(KSH^2+K^2SD)$ & $\mathcal{O}(KSH^2+KS^2D)$ & $\mathcal{O}(S+K)$\\ 
DPRNN \cite{luo2019dual} & $\mathcal{O}(KSH^2)$ & $\mathcal{O}(KSH^2)$ & $\mathcal{O}(S+K)$\\ 
GALR & $\mathcal{O}(KSH^2)$ & $\mathcal{O}(QS^2D)$ & $\mathcal{O}(K)$\\
\specialrule{.16em}{0em}{0em} 
\end{tabular}
\vspace{-0.3cm}
\end{table*}
\subsection{Permutation Invariant Training}
In a standard training framework, given a speech separation model $f_\theta$ with a set of parameters denoted by $\theta$, a loss function $\mathcal{L}(f_\theta(\mathbf{x}), \mathbf{y})$ is used to penalize the divergence between the predicted outputs $f_\theta(\mathbf{x})=\{\hat{\mathbf{s}}_1, ..., \hat{\mathbf{s}}_C\}$ and the clean sources $\mathbf{y}=\{\mathbf{s}_1, ..., \mathbf{s}_C\}$. As an end-to-end network, our proposed GALR model  outputs waveforms of the estimated
clean signals so that we can directly use the scale-invariant source-to-noise ratio (SI-SNR) \cite{luo2018tasnet} as our maximization objection with permutation invariant training (PIT) \cite{yu2017permutation, kolbaek2017multitalker}:
\begin{align}
    \mathcal{L}_\text{SI-SNR}(f_\theta(\mathbf{x}), \mathbf{y})&=-10\log_{10}\frac{\lVert\Pi_{\mathbf{s}}(\hat{\mathbf{s}})\rVert_2^2}{\lVert\hat{\mathbf{s}}-\Pi_{\mathbf{s}}(\hat{\mathbf{s}})\rVert_2^2},
\end{align}
where
$
    \Pi_{\mathbf{a}}(\mathbf{b})=\frac{\mathbf{a}^\top\mathbf{b}}{\lVert\mathbf{a}\rVert_2^2}\mathbf{a}
$ is the projection of $\mathbf{b}$ onto $\mathbf{a}$. SI-SNR has also been used in many end-to-end separation models \cite{luo2018tasnet, luo2019conv, yang2019improved, luo2019dual}.

\section{Evaluation and Analysis}
\label{sec:3}
\subsection{Experimental Setup}

\subsubsection{Data Preparation}
We used three datasets for our experiments: (1) WSJ0-2mix, a two-speaker speech dataset \cite{garofalocontinuous, hershey2016deep} that consists of 30 hours of training, 10 hours of validation, and 5 hours of evaluation data and is widely used as the benchmark in monaural speech separation \cite{luo2019dual, liu2019divide, luo2018tasnet, luo2019conv, wang2018end, zhang2020furcanext, lam2020mixup}, (2) Libri-2mix, a larger two-speaker audio mixture dataset generated from a publicly available English speech corpus Librispeech \cite{panayotov2015librispeech} that contains 982.1 hours of speech from 2484 speakers, and (3) WSJ0-music, a speech-music mixture audio dataset generated in \cite{lam2020mixup}. All mixture audios were simulated by randomly combining utterances from different speakers or music clips at a sampling rate of 8 kHz with SNRs between 0 dB and 5 dB.

\subsubsection{Model Setup}
In our implementation, we used the setting of encoder-decoder modules in \cite{luo2018tasnet, luo2019conv} and the segmentation module described in \cite{luo2019dual}. In the middle part of the separation network, 6 consecutive blocks were used to model local and global sequences, i.e., $N=6$. We fixed the number of hidden nodes ($H$) in Bi-LSTM to 128 as in \cite{luo2019dual}. The multi-head attention based global model was made of 8 heads, i.e., $J=8$. In each attention layer, the dropout rate was set to 0.1. Regarding other model hyperparameters, we varied the number of filters ($D$), the window length ($M$), and the segment size ($K$) as shown in Table \ref{tab:2}. Notably, when we set $D=64$ as in \cite{luo2018tasnet, luo2019conv} the model size of GALR is much smaller than that in the previous works, therefore, we also tried $D=128$ to obtain a comparable model size with DPRNN. Meanwhile, we implemented the most recently proposed DPTNet \cite{chen2020dual} as another reference.
\par
It is worthwhile to explain why we omitted the configuration of $M=2$ and $K=250$, which corresponds to the highest SI-SNRi score in \cite{luo2019dual}. Although the authors in \cite{luo2019dual} reported that setting shorter window length leads to better SI-SNRi performance, the associated computational burden was not disclosed. Given a limited GPU memory, halving the window length resorts to halving the batch size and doubling the training time. We tried to run the highest scores under the setting of $M=2$ for both DPRNN and GALR. However, building such systems in a small dataset like WSJ0-2mix costs more than ten days of training. For a realistic industrial development, we generally need to tackle 10-100 times larger datasets. Therefore, the corresponding training efficiency is unacceptable for most realistic scenarios, not to mention that the high latency at inference time is also problematic for system deployment.

\subsubsection{Training Details}
All the models were trained on 8 NVIDIA Tesla M40 GPU devices using PyTorch \cite{paszke2017automatic} for fair comparisons. We found that the performances of all separation models deteriorated as we used 8 GPUs in place of 1 GPU. A similar observation has been obtained in \cite{su2015experiments, chen2016scalable}, where this multi-GPU training approach is termed model averaging. Note that it is impractical to use a single GPU for model training with its unacceptably long training time. For a fair part-to-part comparison, we report the performances of GALR under the same 8-GPU training condition, though the SI-SNRi can be further improved in the case of using fewer GPUs.
\par
For the benchmark WSJ0-2mix separation task, we referred to the training protocol in \cite{luo2019dual}, where clipped 4-second waveforms were used for permutation invariant training \cite{yu2017permutation} to minimize pairwise SI-SNR loss \cite{luo2018tasnet}. Concerning optimization, we used Adam \cite{kingma2014adam} optimizer with an initial learning rate of $1e^{-3}$ and a weight decaying rate of $1e^{-6}$. The learning rate was exponentially decayed at a rate of 0.96 for every two epochs. The training was considered converged when no lower validation loss can be observed in 10 consecutive epochs. A gradient clipping method was used to ensure the maximum l2-norm of each gradient is less than 5. All models were assessed in terms of SI-SNRi and SDRi \cite{le2019sdr}.

\subsection{Performance Analysis}

\subsubsection{Investigation of GALR Optimality}
 \label{sec:4.2.1}

\begin{table}[h!]
\centering
\caption{SI-SNRi results of WSJ0-2mix when permuting Bi-LSTM and attention model in local and global modeling}
\vspace{-0.1cm}
\label{tab:1}
\begin{tabular}{c|cc}
\specialrule{.16em}{0em}{0em} 
\textbf{Approach} & \textbf{Local Bi-LSTM} & \textbf{Local Attention} \\
\hline
\textbf{Global Bi-LSTM} & 15.9 & 12.3  \\
\textbf{Global Attention} & \textbf{17.0} & 14.6 \\
\specialrule{.16em}{0em}{0em} 
\end{tabular}
\end{table}

First and foremost, we experimented on WSJ0-2mix to validate our hypothesis that the proposed GALR architecture is the optimal choice amongst while permuting recurrent and attention models for local and global sequence modeling.
In this experiment, we chose the bi-directional LSTM as a representative recurrent model. As shown in Table \ref{tab:1}, we obtained 4 distinctive SI-SNRi scores from 4 kinds of TasNet systems. Interestingly, we found two consistent patterns: (1) in local modeling, the recurrent model performed better than the attention model; (2) in global modeling, the attention model performed better than the recurrent model. Overall, the proposed GALR architecture (bottom-left) gave the best performance among the four architectures, which matches our expectations.

\subsubsection{Time Complexity Analysis}
Next, we compared the algorithmic complexities of all three models in Table \ref{tab:0}, where we reported dual-path processing complexities and the maximum path length (MPL) \cite{vaswani2017attention} that was needed to connect any two positions in the signal sequence in Big-O notation. Considering the global-path processing complexity, the cost of DPTNet was about the sum of the costs of both DPRNN and GALR. Notably, for very long input sequences, e.g., in the case of small window length, we needed to use a larger $K$ to improve the computational performance of both models. By introducing low-dimensional mapping with $Q<<K$, we found that GALR could significantly relieve the computational burden carried by a large $K$, as reported in terms of actual FLOPs in Table \ref{tab:2}. Besides, with regard to MPL, amongst the three models, only GALR managed to connect all positions with $\mathcal{O}(K)$, whereas both DPRNN and DPTNet required $\mathcal{O}(S+K)$ sequential operations. As discussed in \cite{hochreiter2001gradient}, the shorter the MPL, the easier it was to learn long-term dependencies, which echoed the theoretical advantage of GALR.

\subsubsection{Visualizing Global Attention with Multiple Heads}
\begin{figure}[th!]
 \centering
    \includegraphics[width=\linewidth]{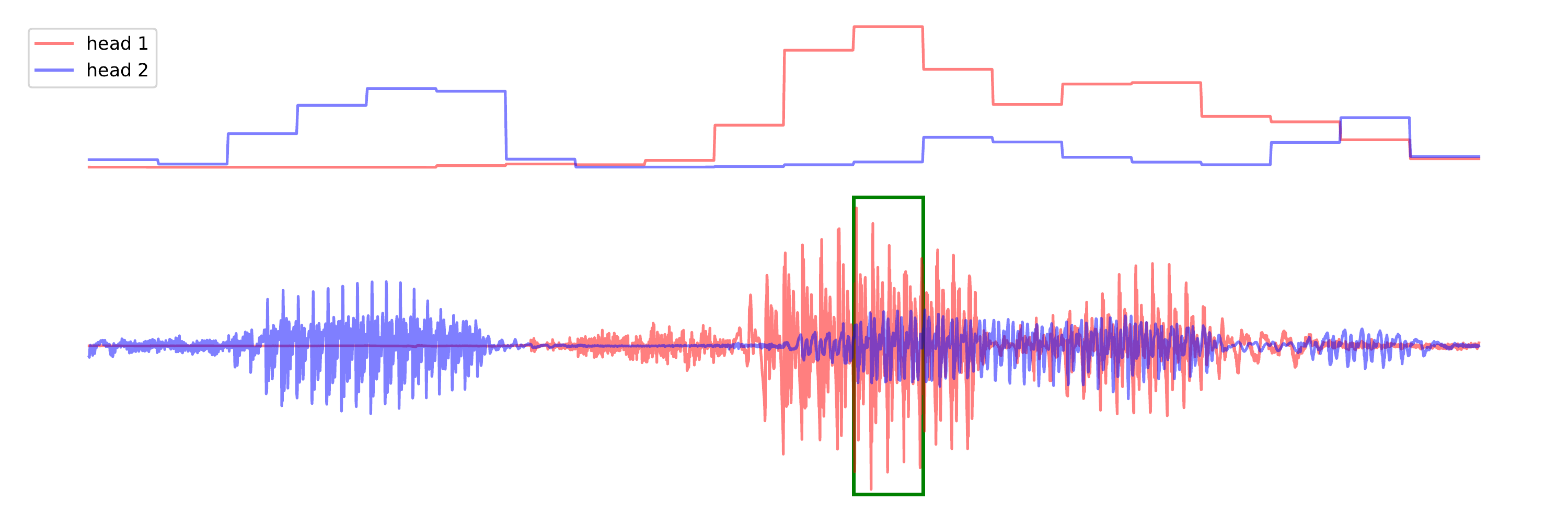}
\caption{An example given by a GALR model trained on WSJ0-2mix. The clean speech signals unseen by the model are shown in red and blue. The two graphs above the signals denotes the softmax values of two selected heads attending on the target segment in green frame.}
\label{fig:attn}
\vspace{-0.2cm}
\end{figure}
We were also interested in understanding how the multi-head self-attention works in speech separation. To reason about the physical meaning of our global attention mechanism, we examined the values of the softmax matrices defined in Eq. \ref{eq:5} computed in different heads. The softmax values were plotted against the time axis with respect to the source signals, as shown in Fig. \ref{fig:attn}. Interestingly, we observed two heads where the attention values were related to the energies of the two clean speeches. This explained how multi-head self-attention mechanism worked in GALR -- the attention matrices of different heads were combined over stacks of GALR blocks to output an easily separable representation for the mixture signal. This attention behavior was also mimicking how humans conceptually following the speech of one talker with regards to its volume.

\begin{table*}[t]
\centering
\caption{Performances of DPRNN and our proposed GALR in WSJ0-2mix test set with different configurations.}
\vspace{-0.1cm}
\label{tab:2}
\begin{tabular}{c|cccc|c|cc|cc}
\specialrule{.16em}{0em}{0em} 
\textbf{TasNet}  & \textbf{$D$} & \textbf{$M$} & \textbf{$K$}& \textbf{$Q$}  & \textbf{Size} & \textbf{SI-SNRi}  & \textbf{SDRi} & \textbf{Runtime Memory} & \textbf{GFLOPs} \\
\hline
DPTNet \cite{chen2020dual} & 64 & 16 & 100& -  & 2.8M& 15.5 & 15.7 & 419 MiB&12.6\\ 
\hline
\multirow{3}{*}{DPRNN \cite{luo2019dual}}& 64 & 16 & 100& \multirow{3}{*}{-} & \multirow{3}{*}{2.6M} & 15.9 & 16.2 & 231 MiB & 10.7\\ 
&64& 8 & 150 &&& 17.0 & 17.3 & 456 MiB & 22.2 \\ 
&64& 4 & 200 &&& 17.9 & 18.1 &929 MiB & 42.3\\ 
\hline
\multirow{6}{*}{GALR} & 64 &16 & 100 & 32& \multirow{3}{*}{\textbf{1.5M}} & \textbf{16.2} &  \textbf{16.5} & \textbf{161 MiB} &\textbf{5.6}\\
 &64& 8 & 150 & 16&&  \textbf{17.1} &  \textbf{17.4} &  \bf{309 MiB} &\bf{11.5} \\
&64& 4 & 200& 8 & & 17.7&  17.9 &\bf{594 MiB} &\bf{21.4}\\ \cline{2-10}
 &128& 16 & 100 & 32 & \multirow{3}{*}{\textbf{2.3M}} & \textbf{17.0}  &  \textbf{17.3} & \textbf{186 MiB} &\textbf{8.3}\\
 &128& 8 & 150 & 16&& \textbf{18.7} &  \textbf{18.9}&  \bf{363 MiB} &\bf{16.5} \\
&128& 4 & 200& 8  && \textbf{20.3} &  \textbf{20.5} &\bf{730 MiB} &\bf{30.8}\\
\specialrule{.16em}{0em}{0em} 
\end{tabular}
\vspace{-0.3cm}
\end{table*}

\subsubsection{Performances in Benchmark WSJ0-2mix}
\label{sec:4.2.3}

As for the architecture for TasNet, we compared the result of our GALR model to the state-of-the-art DPRNN \cite{luo2019dual} and DPTNet \cite{chen2020dual}. Considering only the lightest model with $M=16$, we found that DPTNet costs 152\% more time and 225\% more memory than our proposed GALR. Considering this remarkable surge of memory consumption, we noted that DPTNet might be not practically preferable for most realistic industrial tasks; therefore we only compared GALR with DPRNN in the following tasks.
We replicated the experiment in \cite{luo2019dual} with the same configurations of window length and segment size. The results are shown in Table \ref{tab:2}. Besides the standard separation measure SI-SNRi, we also analyzed the runtime cost of each model for processing a 1s mixture input in terms of memory measured in GPU and floating-point operations (FLOPs) approximated with a third-party module,\footnote{https://github.com/sovrasov/flops-counter.pytorch} which represents the model efficiency. On the one hand, the GALR with a model size comparable to DPRNN consistently gave superior SI-SNRi performances over DPRNN under the same configurations of window length and segment size. On the other hand, the smaller GALR attained comparable or better separation performances while requiring only 57.3\% parameters and reducing up to 36.1\% runtime memory and 49.4\% computational operations.

\subsubsection{Performances in Large-scale Libri-2mix}

\begin{table}[h!]
\centering
\caption{Performances of DPRNN and our proposed GALR in Libri-2mix}
\label{tab:3}
\vspace{-0.1cm}
\begin{tabular}{c|c|c|c}
\specialrule{.16em}{0em}{0em} 
\textbf{TasNet} & \textbf{Size}  & \textbf{SI-SNRi}  & \textbf{SDRi} \\
\hline
DPRNN \cite{luo2019dual} & 2.6M & 12.0 & 12.5\\
\hline
GALR& \textbf{2.3M} & \textbf{12.2}& \textbf{12.7}\\
\specialrule{.16em}{0em}{0em} 
\end{tabular}
\end{table}

To validate the consistency of GALR's improvements over DPRNN, we experimented on a larger dataset Libri-2mix. In Libri-2mix, we split the Librispeech \cite{panayotov2015librispeech} corpus into (1) a training set containing 12055 utterances (5 utterances per speaker) drawn from 2411 speakers, (2) a validation set containing another 7233 utterances (5 utterances per speaker) drawn from the same 2411 speakers, and 2411 speakers (3) a test set containing 4380 utterances (60 utterances per speaker) drawn from 73 speakers. Note that the separation task in Libri-2mix was much more challenging than that in WSJ0-2mix because of not only increasing training speakers from 101 to 2411 but also the limitation of only five sampled utterances per speaker for training. Despite the increased separation difficulty, we conceived that the separation scenario became more realistic as it was often hard to collect many clean utterances from the user in real-world applications.
\par
In this large-scale separation dataset, to strike a balance between the training speed and the separation performance, we decided to use a configuration of $M=8$ and $K=150$ to train both DPRNN and GALR. For a fair comparison in terms of the number of parameters, we only trained the GALR with 2.3M parameters to be comparable to DPRNN. The separation results of DPRNN and GALR were summarized and shown in Table \ref{tab:3}. We observed that GALR maintained its advantage in speed and memory over DPRNN and meanwhile achieved 0.2dB absolute improvements in SI-SNRi and SDRi. The consistent results suggest that GALR keeps performing low-cost separation without sacrificing separation efficacy.

\subsubsection{Performances in Separating Speech-Music Mixture}

\begin{table}[h!]
\centering
\caption{Performances of DPRNN and our proposed GALR in WSJ0-music}
\label{tab:4}
\vspace{-0.1cm}
\begin{tabular}{c|c|c|c}
\specialrule{.16em}{0em}{0em} 
\textbf{TasNet} & \textbf{Size}  & \textbf{SI-SNRi}  & \textbf{SDRi} \\
\hline
DPRNN \cite{luo2019dual} & 2.6M & 14.5 & 14.8\\
\hline
GALR& \textbf{2.3M} & \textbf{15.9}& \textbf{16.2}\\
\specialrule{.16em}{0em}{0em} 
\end{tabular}
\end{table}

Besides two-speaker speech separation tasks, we were also interested in separating speech from a speech-music mixture. In this paper, we simulated the speech-music mixture using the speech from the WSJ0-2mix corpus and the music clips in \cite{lam2020mixup}. The results were shown in Table \ref{tab:4}. Comparing to the two-speaker separation, we found that GALR obtained a even greater superiority over DPRNN in the speech-music scenario, which is also very common in real-world conversation scenarios. In particular, there is a growing demand in the industry where the speech-music separation becomes critical for the deployment of many real-world signal processing systems, e.g., micro-video automatic captioning. Therefore, the consistent and larger superiority of GALR over DPRNN in the speech-music separation task is valuable for both conventional and emerging application deployments.

\section{Conclusions}
\label{sec:4}
This paper introduces a novel architecture -- globally attentive locally recurrent network (GALR), which combines the advantages of recurrent networks and attention mechanisms for effective, low-cost time-domain signal processing. We provide empirical evidences that the self-attention mechanism is a better candidate for modeling the long-range context sequence than the RNNs. Results across three different datasets also suggest the superiority of attention models over recurrent models in modeling global sequences, which leads to greater modeling power, reduced model size, and less runtime memory.
We believe that a compact, low-cost, and effective separation system is more practical to the industry and will empower wider applications of speech separation for robust signal processing.

\bibliographystyle{IEEEbib}
\bibliography{strings,refs}

\end{document}